\documentclass
[lenghhestimate, tighten, amssymb, aps, floatfix, pra, twocolumn, tightenlines,superscriptaddress]{revtex4-1}
\pdfoutput=1
\usepackage{graphics}
\usepackage{graphicx,color}
\usepackage{subcaption} 
\usepackage{float}
\usepackage[ansinew]{inputenc}
\usepackage{graphicx}
\usepackage{amsmath}
\usepackage{amssymb}
\usepackage{dsfont} 
\usepackage{hyphenat}
\usepackage[percent]{overpic}
\usepackage{xcolor}
\usepackage{appendix}
\usepackage{bbold}

\begin{document}

\title{Cryogenic operation of silicon photonic modulators based on DC Kerr effect}
\author{Uttara Chakraborty}
\email{uttara@mit.edu}
\affiliation{Research Laboratory of Electronics, Massachusetts Institute of Technology, Cambridge, Massachusetts 02139, USA}
\author{Jacques Carolan}
\affiliation{Research Laboratory of Electronics, Massachusetts Institute of Technology, Cambridge, Massachusetts 02139, USA}
\author{Genevieve Clark}
\affiliation{The MITRE Corporation, Bedford, Massachusetts 01730, USA}
\author{Darius Bunandar}
\affiliation{Research Laboratory of Electronics, Massachusetts Institute of Technology, Cambridge, Massachusetts 02139, USA}
\author{Jelena Notaros}
\affiliation{Research Laboratory of Electronics, Massachusetts Institute of Technology, Cambridge, Massachusetts 02139, USA}
\author{Michael R. Watts}
\affiliation{Research Laboratory of Electronics, Massachusetts Institute of Technology, Cambridge, Massachusetts 02139, USA}
\author{Dirk R. Englund}
\affiliation{Research Laboratory of Electronics, Massachusetts Institute of Technology, Cambridge, Massachusetts 02139, USA}

\date{\today}

\noindent
\begin{abstract}

Reliable operation of photonic integrated circuits at cryogenic temperatures would enable new capabilities for emerging computing platforms, such as quantum technologies and low-power cryogenic computing.
The silicon-on-insulator platform is a highly promising approach to developing large-scale photonic integrated circuits due to its exceptional manufacturability, CMOS compatibility and high component density.
Fast, efficient and low-loss modulation at cryogenic temperatures in silicon, however, remains an outstanding challenge, particularly without the addition of exotic nonlinear optical materials.
In this paper, we demonstrate DC-Kerr-effect-based modulation at a temperature of 5~K at GHz speeds, in a silicon photonic device fabricated exclusively within a CMOS process.
This work opens up the path for the integration of DC Kerr modulators in large-scale photonic integrated circuits for emerging cryogenic classical and quantum computing applications.

\end{abstract}
\maketitle

\section{Introduction} 
\label{sec:introduction}


Photonic integrated circuits (PICs) have emerged as a promising platform for a wide range of integrated optical computing and signal processing applications. PICs are a highly attractive candidate for quantum technologies due to the ease with which they can generate, manipulate and detect quantum states of light in a compact architecture \cite{wang2019integrated}. %
These systems can now be accessed through commercial photonics foundries \cite{hochberg2010towards}, paving the way for a new paradigm in large-scale photonic quantum information processing \cite{Carolan:2015vga, wang2019integrated} and quantum key distribution \cite{sibson2017chip, bunandar2018metropolitan}. 
The CMOS-compatibility and large refractive index contrast of the silicon-on-insulator platform makes it particularly well suited for large-scale quantum photonics \cite{Silverstone:2016gha, Rudolph:2017du} and optical data transmission. 

However, integration of silicon photonics with key quantum componentry --- including single photon emitters \cite{kim2020hybrid, elshaari2020hybrid}, quantum memories \cite{dibos2018atomic} and single photon detectors \cite{Najafi:2014ey, shibata2019waveguide} --- will likely require operation at cryogenic ($<10$~K) temperatures. Silicon photonics provides an efficient platform for electro-optic interconnects, which can reduce passive heat loading and provide a high-speed, low-loss solution to limitations on electrical connections in cryogenically-cooled circuits \cite{youssefi2020cryogenic}. Low-power cryogenic systems for classical computing, such as superconducting optoelectronic circuits \cite{shainline2018circuit}, must also be operated at $<10$~K temperatures \cite{holmes2013energy}.

The major challenge in the operation of silicon quantum photonics at cryogenic temperatures is the absence of fast, low-loss, phase-only modulation technology, which is required for active multiplexing \cite{gimeno2017relative} and high-fidelity quantum gates \cite{Laing:2010fk}.
To date, silicon photonic modulators typically rely on the plasma dispersion effect \cite{Reed:2010im, timurdogan2014ultralow, perez2015comparison} or the (typically slow) thermo-optic effect \cite{Harris:2014kz}.
At cryogenic temperatures, thermo-optic modulators suffer from a decrease in the thermo-optic coefficient of silicon by four orders of magnitude \cite{Komma:2012dj}. Plasma dispersion modulators can suffer from large carrier-induced losses and can be bandwidth-limited by low-temperature carrier freeze-out \cite{gehl2017operation, pires1990carrier}.

While silicon's intrinsic third order $\chi^{(3)}$ nonlinearity makes it an excellent candidate for nonlinear processes such as four-wave mixing \cite{faruque2018chip, carolan2019scalable}, silicon's crystal centro-symmetry precludes the existence of a native second-order $\chi^{(2)}$ nonlinearity. Thus, unlike non-centro-symmetric materials such as lithium niobate (\text{LiNbO$_{3}$}) and gallium arsenide (GaAs), silicon cannot be a natural candidate for Pockels-effect-based electro-optic modulators.
One approach is to integrate materials that possess a strong second-order nonlinearity such as \text{LiNbO$_{3}$} \cite{he2019high} or barium titanate (\text{BaTiO$_{3}$}) \cite{eltes2019integrated} into the silicon photonics platform. 
Such hybrid integration techniques, however, typically require non-standard fabrication procedures and complex material compatibility, thus motivating the need for a ``zero-change'' phase-only modulator in silicon.

The absence of an intrinsic $\chi^{(2)}$ in silicon had, until recently, precluded the development of phase-only modulators in pure silicon. While some prior experiments \cite{jacobsen2006strained, chmielak2011pockels} have suggested that a large second-order nonlinearity can be induced by straining silicon waveguides, more recent studies \cite{olivares2017recent, castellan2019origin} have deduced that the reported second-order nonlinearities were overestimates, and that the observed nonlinear effects arose from interfacial charge defects rather than strain as originally believed.  Alternatively, it has now been shown \cite{Timurdogan:2017jg, castellan2019field} that a second-order nonlinearity can be generated in silicon by means of an applied DC electric field, producing an effective $\chi^{(2)}$ from the third order nonlinear susceptibility $\chi^{(3)}$. This approach leads to efficient electric-field-induced second harmonic (EFISH) generation in silicon as well as a large electro-optic DC Kerr effect that can be leveraged in optical phase shifters. 

The index shift in silicon due to an applied DC field is derived using the following approach, where  $E_{0}$ denotes the magnitude of the static applied electric field and $E_{\omega}$ the amplitude of the optical mode field with frequency $\omega$.  
Since $\chi^{(2)}$ = 0 in silicon, the nonlinear polarization \cite{boyd2019nonlinear} is given by 
\begin{equation}
\ P = \epsilon_{0}(\chi^{(1)}(E_{0}+E_{\omega}e^{j \omega t})+\chi^{(3)}(E_{0}+E_{\omega}e^{j \omega t})^{3} + ...)
\end{equation}
Retaining only the terms proportional to $E_{\omega}$, we have
\begin{equation}
\ P = \epsilon_{0}(\chi^{(1)}+3\chi^{(3)}E_{0}^{2})E_{\omega}e^{j \omega t}+...
\end{equation}
where  $\chi^{(1)}+3\chi^{(3)}E_{0}^{2}$ is the effective relative permittivity $\epsilon_{r}$.
As such, the change in relative permittivity with an applied DC field $E_{0}$ is given by 
\begin{equation}
\Delta\epsilon_{r}= 3 \chi^{(3)}E_{0}^{2}. 
\end{equation}
Since the refractive index $n$ equals $\sqrt{\epsilon_{r}}$,
we obtain the electric-field-induced refractive index shift as
\begin{equation}
\Delta n = \frac{3 \chi^{(3)}E_{0}^{2}}{2n}. 
\end{equation} 

The performance of DC-Kerr-effect-based modulators incorporated into Mach-Zehnder interferometers (MZIs) has been demonstrated at room temperature \cite{Timurdogan:2017jg}. The use of such DC-Kerr-effect-based modulators at cryogenic temperatures is highly desirable in photonic integrated circuits for classical and quantum information processing. Based on Equation (4), given that the thermo-optic decrease in the refractive index of silicon between room and cryogenic temperatures is negligibly small (on the order of $10^{-2}$ \cite{Komma:2012dj, frey2006temperature}), the refractive index shift at cryogenic temperatures is expected to be of the same order of magnitude as that at room temperature for the same externally-applied DC electric field.
In the present work, we demonstrate the cryogenic operation of integrated DC-Kerr-based silicon modulators, showing that the effective refractive index shifts observed at a 
temperature of 5~K are indeed comparable to those observed at room temperature.

\begin{figure}[!tb]
\centering
\includegraphics[width=\linewidth]{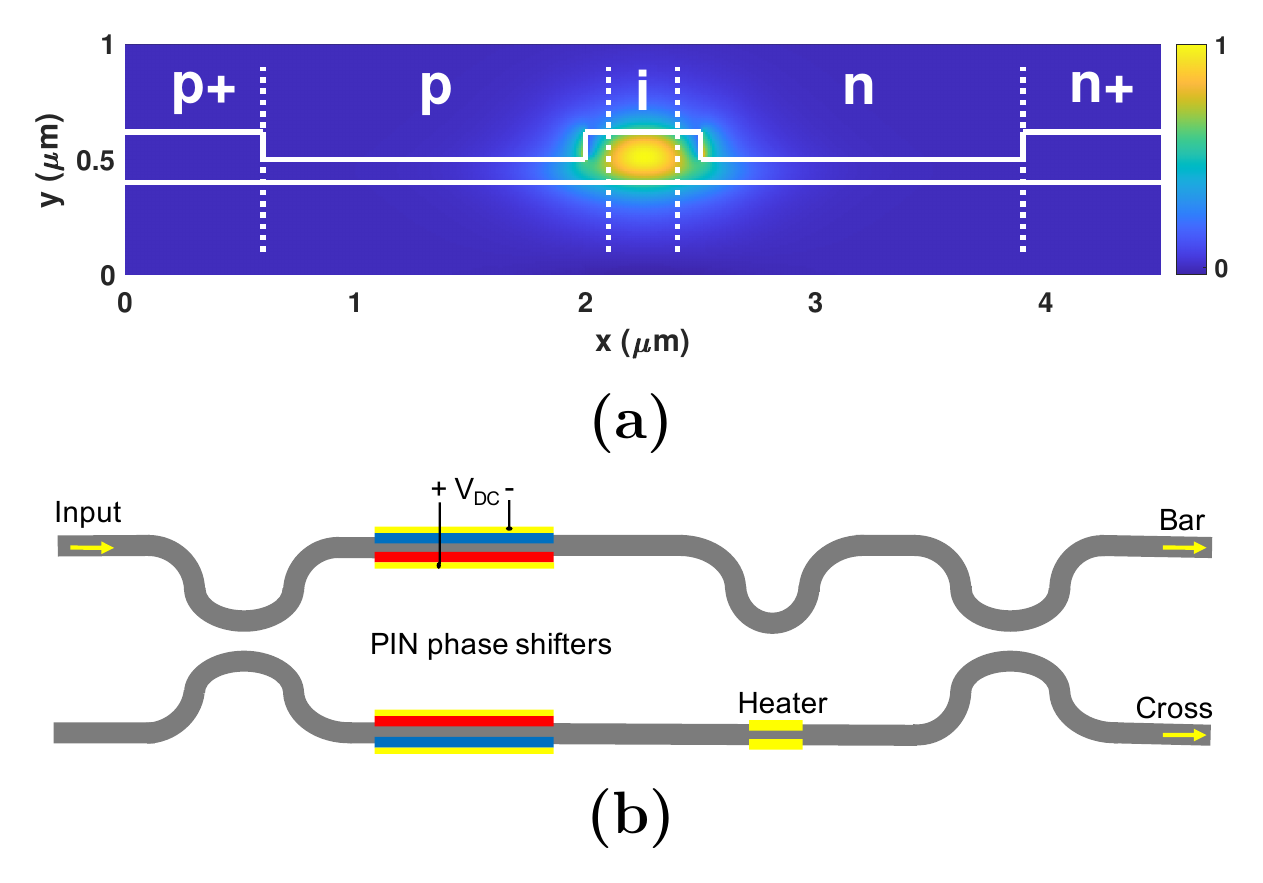}
\caption{\small \textbf{Integrated PIN modulator device.} 
\textbf{(a)} PIN modulator cross section with normalized fundamental TE mode profile ($E_{x}$). \textbf{(b)} Unbalanced Mach-Zehnder interferometer with an integrated PIN phase shifter in each arm.}
\label{fig:device}
\end{figure}

\section{Device structure}
\label{sec:Device structure}

Our test structures were two MZIs with input and output adiabatic couplers, 4.5-mm-long integrated PIN junction modulators in both arms, and a path length difference of 37.4 $\mu$m between the arms (Fig.~\ref{fig:device}). (One arm of each MZI also had an embedded thermo-optic phase shifter which was not used in this experiment.) The devices were fabricated in a CMOS foundry (at SUNY Polytechnic Institute) using silicon-on-insulator wafers following the design in \cite{Timurdogan:2017jg}. The modulators consisted of 500 nm $\times$ 220 nm silicon ridge waveguides with embedded PIN junctions.  The widths of the intrinsic silicon regions of the waveguide core of the devices were 300 nm and 400 nm, respectively. The p and n regions were formed by boron difluoride (\text{BF$_{2}$}) and arsenic (As) implants with a target doping concentration of $10^{18}$ cm$^{-3}$, and the p+ and n+ regions by \text{BF$_{2}$} and phosphorus (P) implants with a target doping concentration of $10^{20}$ cm$^{-3}$. Tungsten vias contacting the n+ and p+ regions provided electrical connections to copper routing layers and on-chip copper contact pads. We present the experimental results from the 300 nm intrinsic silicon device in the main body of the paper and the results from the 400 nm intrinsic device in Appendix~\ref{sec:appendix-400}.

\section{Experimental Procedure}
\label{sec:Experimental Procedure}
We affixed the chip with the test structures to a copper mount with conductive silver paint (Ted Pella) and wire-bonded it to a high-frequency PCB. We mounted the package in a closed-cycle cryostat (Montana Instruments) and connected the PCB to the cryostat's RF feed-throughs. Using piezo-driven positioners, we edge-coupled lensed single-mode fibers to the input and output adiabatic couplers of the Mach-Zehnder devices (with a measured loss of $\sim$6 dB per facet). Appendix~\ref{sec:appendix-config} shows the experimental setup with the packaged chip mounted in the cryostat. 

\begin{figure}  
\centering
\includegraphics[trim=0 0 0 0, clip, width=1.0\linewidth]{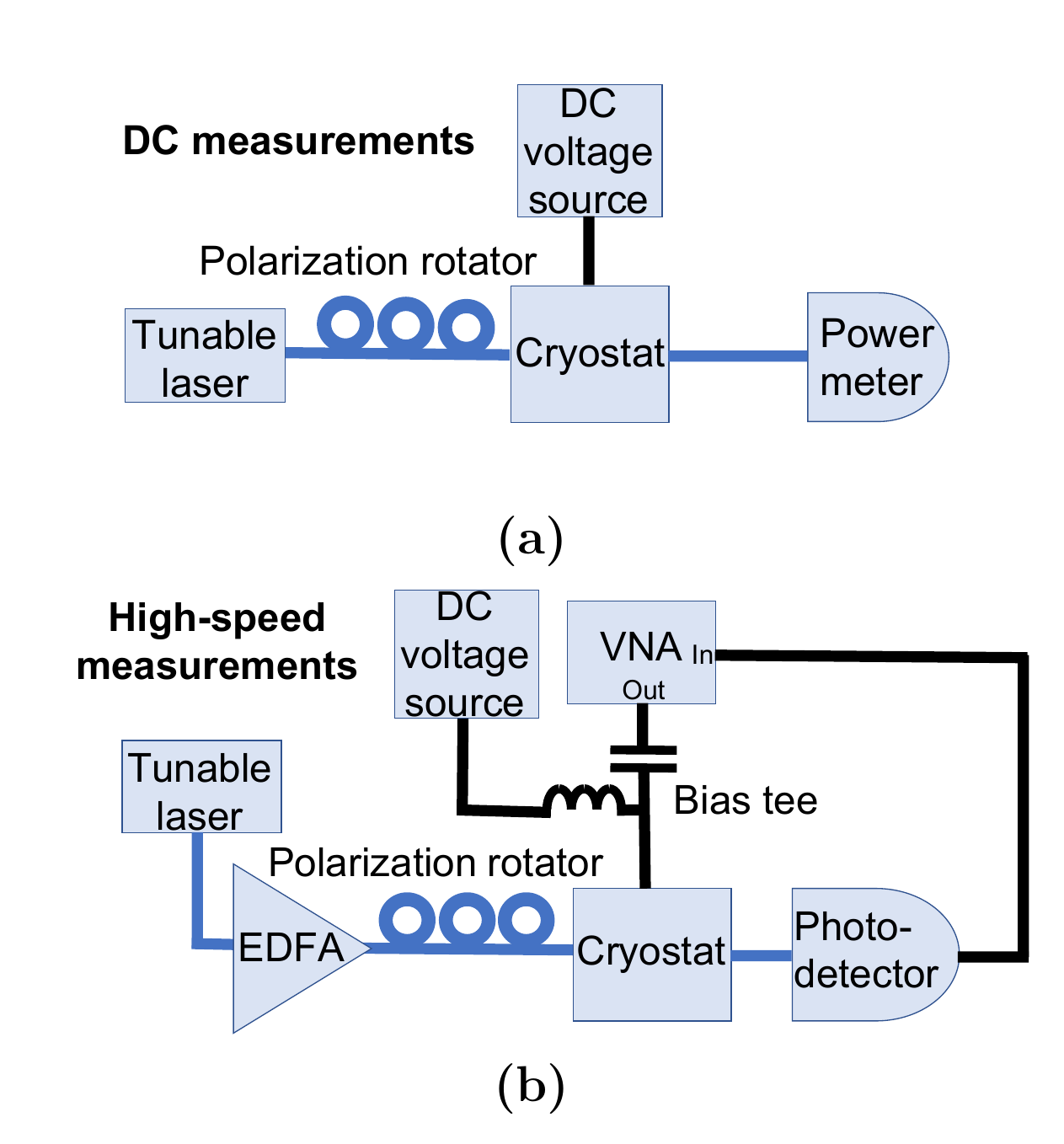}
\caption{\small \textbf{Experimental configurations.} Optical routing shown in blue and electrical in black. \textbf{(a)} DC measurements:  Tunable laser input is polarization-matched to waveguide mode and Mach-Zehnder output transmission is detected with a telecom photodetector. DC reverse bias is applied to one device at a time. \textbf{(b)} RF measurements: Tunable laser input is polarization-matched to waveguide mode and amplified before being routed to Mach-Zehnder input. Through a bias tee, a DC bias is applied to operate each device at its quadrature point, and an AC small signal is simultaneously applied from the VNA. The Mach-Zehnder optical output is routed to a high-speed photodetector and the photodetector electrical output is fed back to the VNA.}
\label{fig:configurations}
\end{figure}

Fig.~\ref{fig:configurations} shows the two experimental configurations for measurements conducted with DC reverse bias only and DC bias with AC small signal, respectively. We conducted transmission measurements at room temperature and at 5~K. An Agilent tunable telecom laser (1 mW) was coupled through the input lensed fiber onto the on-chip waveguide, and polarization rotators were used to match the input to the fundamental waveguide TE mode as closely as possible. For DC measurements, we applied voltages (with an Agilent DC power supply) to reverse bias the modulator on one arm of each device at a time, and measured (with an Agilent telecom power meter) the transmitted power from one output arm of the adiabatic coupler at a time (Fig.~\ref{fig:configurations}a). 

To conduct high-frequency measurements for determining the Mach-Zehnder bandwidths, we amplified the input laser with an Oprel erbium-doped fiber amplifier (EDFA) to provide $\sim$5 mW of on-chip power, and applied a 0.1 V AC small signal with frequencies up to 4 GHz using a Keysight vector network analyzer (VNA) (Fig.~\ref{fig:configurations}b). The devices were biased at their quadrature points using the DC power supply. The transmitted optical signal was routed to a Discovery Semiconductors high-speed photodetector, the electrical output of which was in turn routed to the VNA to capture the RF bandwidth of the devices.

\begin{figure*}[!tb]
\centering
\includegraphics[trim=0 0 0 0, clip, height=4.5in]{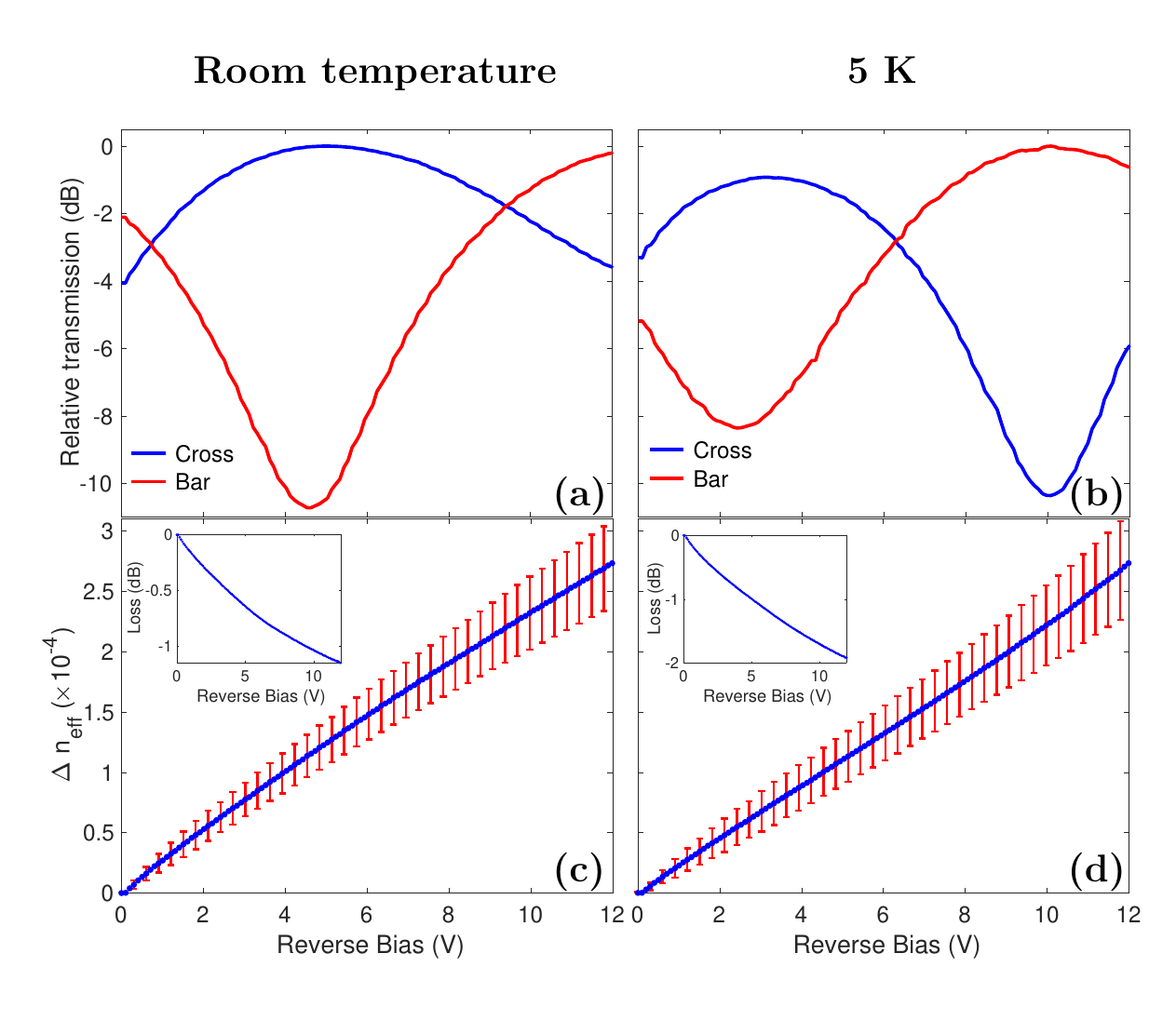}
\caption{\small \textbf{Experimental results for 300 nm intrinsic silicon device.} 
The left column shows results at room temperature, and the right column results at 5K. \textbf{(a)} Optical output from both arms ($\lambda$ = 1548 nm) as a function of applied DC reverse bias at room temperature. \textbf{(b)} Optical output from both arms ($\lambda$ = 1548 nm) as a function of applied DC reverse bias at 5~K. \textbf{(c)} Effective index shift as a function of applied DC reverse bias at room temperature, averaged over input wavelengths. The error bars (shown every three voltage points to avoid clutter) indicate variations arising from imperfections in adiabatic coupler splitting at different wavelengths and experimental errors from fiber-to-chip coupling.  \textbf{(d)} Effective index shift as a function of applied DC reverse bias at 5~K. Insets in \textbf{(c)} and \textbf{(d)} show average relative loss as a function of applied reverse bias.}
\label{fig:DCresults}
\end{figure*}

\begin{figure}[t!]
          \centering
          \includegraphics[trim=0 0 0 0, clip, width=1.0\linewidth]{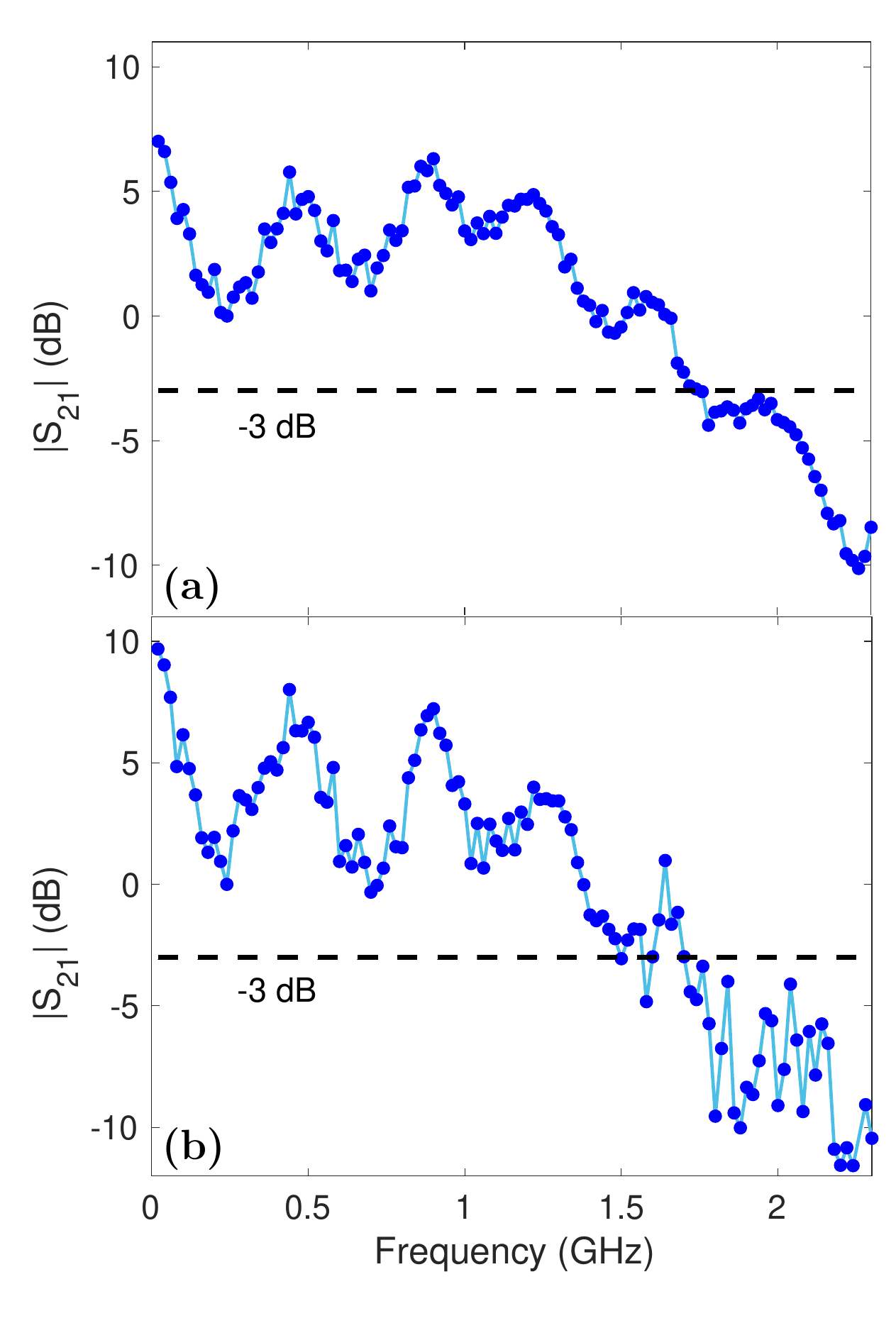}
          \caption{\small \textbf{Frequency response for 300 nm intrinsic silicon device.} \textbf{(a)} Normalized frequency response at room temperature. The dashed line shows the 3 dB cutoff. \textbf{(b)} Normalized frequency response at 5~K.}
          \label{fig:RFresults}
\end{figure}

\section{Results}
\label{sec:Results}
Experimental results for the 300 nm intrinsic silicon device are presented in 
Figs.~\ref{fig:DCresults} and~\ref{fig:RFresults}, and the corresponding results for the 400 nm device in Appendix~\ref{sec:appendix-400}.
Figs.~\ref{fig:DCresults}a and~\ref{fig:DCresults}b 
show the transmitted power from both cross and bar arms at a representative wavelength of 1548 nm for the 300 nm intrinsic silicon MZI at room temperature and at 5~K, respectively. To determine the shift in the effective refractive index at each wavelength as a function of the DC reverse bias, we set the laser input to each wavelength and swept the DC reverse bias on one PIN modulator from 0 V to 12 V for the 300 nm device and 0 V to 17 V for the 400 nm device. The transmission curves showing both a peak and a trough in the applied voltage range were normalized and fitted to a cosine function to extract the differential phase shift as a function of the reverse bias voltage. From this empirically-estimated differential phase shift $\Delta \phi$, we calculated the effective index shift $\Delta n_{eff}$ as $\Delta \phi \times \frac{\lambda}{2\pi L}$, where $\lambda$ is the input wavelength and $L$ the PIN modulator length (4.5 mm).

In Fig.~\ref{fig:DCresults}c, 
we show the average effective index shift obtained from voltage sweeps at wavelengths of 1525-1575 nm for the 300 nm intrinsic silicon device at room temperature. Fig.~\ref{fig:DCresults}d shows the corresponding results at 5~K.  The figure insets show the average relative loss as a function of applied DC reverse bias. With increasing reverse bias, the widening of the depletion region leads to a smaller overlap of the optical mode with free carriers, thus reducing the propagation loss. 
From our results for both the devices, we observe that the magnitudes of the effective index shifts at 5~K remain comparable to those at room temperature, exceeding 2 $\times$ $10^{-4}$ at the upper ends of the applied reverse bias ranges. 
The order of magnitude 
of our experimentally-obtained effective index shifts are in agreement with the simulations presented in Appendix~\ref{sec:appendix-simu}. These experimental results suggest that the DC Kerr effect, as well as the plasma dispersion effect, continues to remain at work down to a temperature of 5~K. 

Fig.~\ref{fig:RFresults} shows the electro-optic frequency response of the 300 nm intrinsic silicon device reverse biased at the quadrature point at room temperature and 5~K, respectively. The 3 dB bandwidths exceed 1.5 GHz at both room temperature and 5~K. Preliminary measurements conducted using an RF probe at room temperature (before the chip was wire-bonded to the PCB) yielded higher RC-limited bandwidths ($\sim$4 GHz), indicating that the chip-PCB packaging ultimately limits the speed of modulation in this experiment. By modifying the electrical packaging in future experiments to overcome the PCB-induced limitation, the devices could be operated at higher speeds up to their full intrinsic bandwidths.

\section{Conclusion and Outlook}
\label{sec:Conclusion}
In this work, we have demonstrated DC-Kerr-effect-based modulation in silicon at a cryogenic temperature of 5~K. 
We studied the electric field-induced refractive index shift in Mach-Zehnder modulators with embedded PIN phase shifters, and showed that effective refractive index shifts
greater than 2 $\times$ 10$^{-4}$ can be achieved at both room temperature and 5~K. The total index shift is the result of a combination of the DC Kerr and plasma dispersion effects. 

Silicon DC Kerr modulators such as the ones demonstrated in this work can be fabricated in commercial silicon photonics foundries without the need for complex fabrication processes involving heterogeneous integration of electro-optic materials \cite{eltes2018first, alexander2018nanophotonic, koeber2015femtojoule}.
However, scaling up photonic quantum technologies to a level where they can tackle practical problems in quantum simulation \cite{AspuruGuzik:2012ho} and machine learning \cite{steinbrecher2019quantum} and ultimately demonstrate a quantum advantage \cite{Aaronson:2011tja, harrow2017quantum}, will likely require the co-integration of photonics with CMOS electronics \cite{atabaki2018integrating, kim2019single}. 
The use of CMOS driving electronics will bring with it new challenges for the large-scale control of DC Kerr modulators. Rigorous engineering of modulator geometries will be essential for maximizing the induced effective $\chi^{(2)}$ by applying transverse electric fields approaching the breakdown field of silicon ($\sim$40 V/$\mu$m) while minimizing losses due to free carrier absorption. 
DC Kerr modulators in other CMOS-compatible platforms, such as silicon nitride (\text{Si$_{3}$N$_{4}$}), could also be explored as a pathway to cryogenic phase modulation in photonic integrated circuits. Recently, the photogalvanic effect in \text{Si$_{3}$N$_{4}$} microring resonators has been shown to build up a DC field which in turn induces a second-order nonlinearity for highly efficient second harmonic generation \cite{lu2020efficient}. A similar photogalvanic approach could potentially enable DC-Kerr-based phase modulation in \text{Si$_{3}$N$_{4}$} or silicon-rich \text{Si$_{3}$N$_{4}$} waveguides at cryogenic temperatures. Our low-temperature, high-speed DC Kerr modulator demonstration now paves the way for the use of such devices in cryogenic quantum and classical computing systems. 

\vspace{0.4cm}
\noindent
{\bf Acknowledgements}:
U.C. is supported by a National Defense Science and Engineering Graduate (NDSEG) Fellowship. J.C. is supported by EU H2020 Marie Sklodowska-Curie Grant Number 751016. Experiments were supported in part by AFOSR grant FA9550-16-1-0391, supervised by G. Pomrenke. Devices were fabricated under the Defense Advanced Research Projects Agency (DARPA) DODOS program (Grant No. HR0011-15-C-0056). This work was supported in part by the MITRE Quantum Moonshot Program. We are grateful to Gerald Gilbert, Michael Fanto, Michael Walsh, Ryan Hamerly, and Carlos Rios Ocampo for helpful discussions and comments. 

\bibliographystyle{apsrev4-1}
\bibliography{jc_bib_uttara}

\clearpage

\appendix

\onecolumngrid

\renewcommand\thefigure{\thesection.\arabic{figure}}

\section{Experimental results of the 400 nm device} 
\label{sec:appendix-400}
\setcounter{figure}{0}
Fig.~\ref{fig:400results} presents the experimental results of the 400 nm device. The magnitude of the overall effective index shift at 5~K remains comparable to that at room temperature.

\begin{figure*}[!h]
\centering
\includegraphics[trim=0 0 0 0, clip, width=1.0\linewidth]{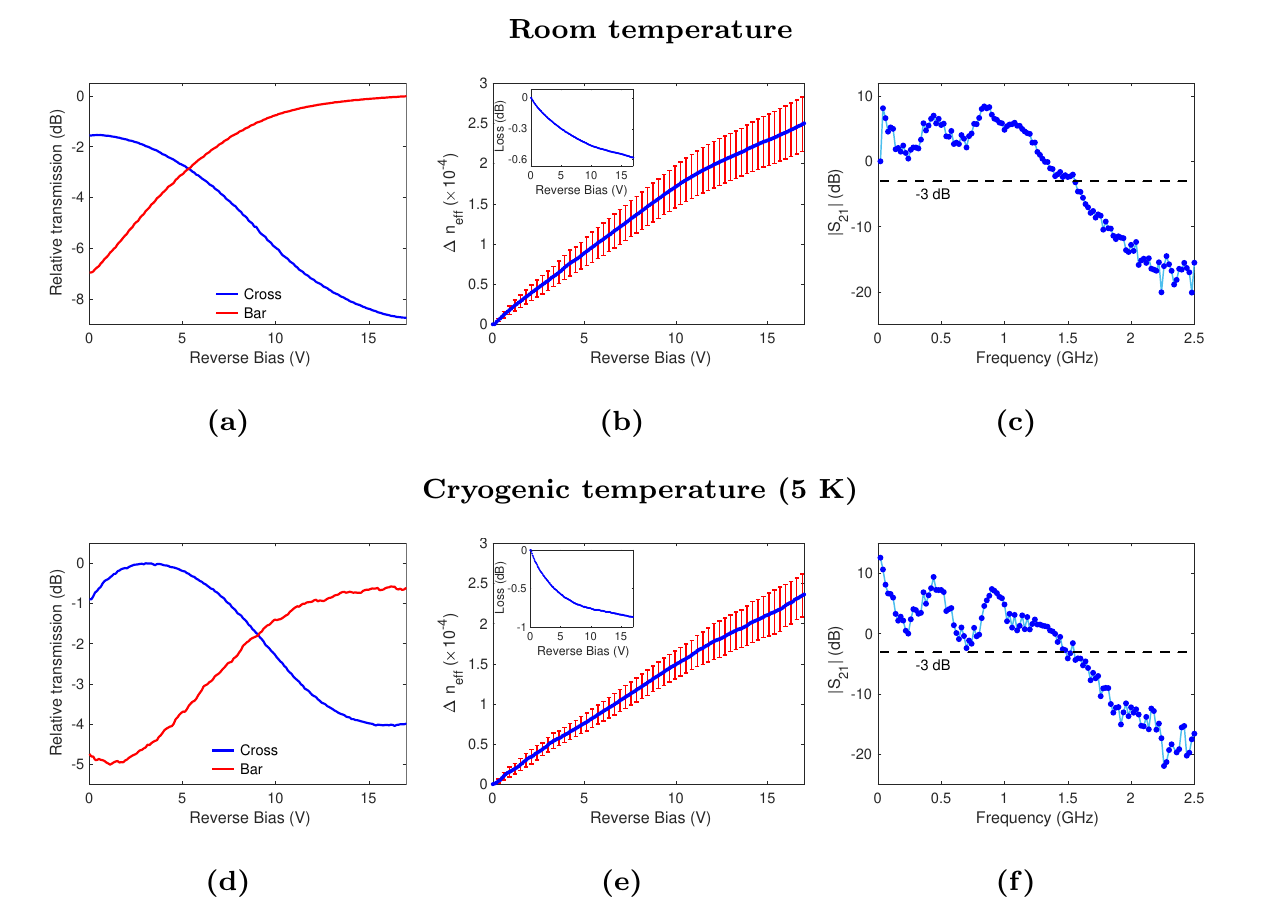}
\caption{\small \textbf{Experimental results for 400 nm intrinsic silicon device.} 
The top row shows results at room temperature, and the bottom row results at 5~K. \textbf{(a)} Optical output from both arms ($\lambda$~=~1548 nm) as a function of applied DC reverse bias at room temperature. \textbf{(b)} Effective index shift as a function of applied DC reverse bias at room temperature, averaged over input wavelengths. The error bars (shown every three voltage points) indicate variations arising from imperfections in adiabatic coupler splitting at different wavelengths and experimental errors from fiber-to-chip coupling.   \textbf{(c)} Normalized frequency response at room temperature. The dashed line shows the 3 dB cutoff. \textbf{(d)} Optical output from both arms ($\lambda$~=~1548~nm) as a function of applied DC reverse bias at 5~K. \textbf{(e)} Effective index shift as a function of applied DC reverse bias at 5~K. \textbf{(f)} Normalized frequency response at 5~K. Insets in \textbf{(b)} and \textbf{(e)} show average relative loss as a function of applied reverse bias.}

\label{fig:400results}
\end{figure*}

\section{Experimental configuration}  
\label{sec:appendix-config}
\setcounter{figure}{0}

\begin{figure}[t!]
\includegraphics[width=\linewidth]{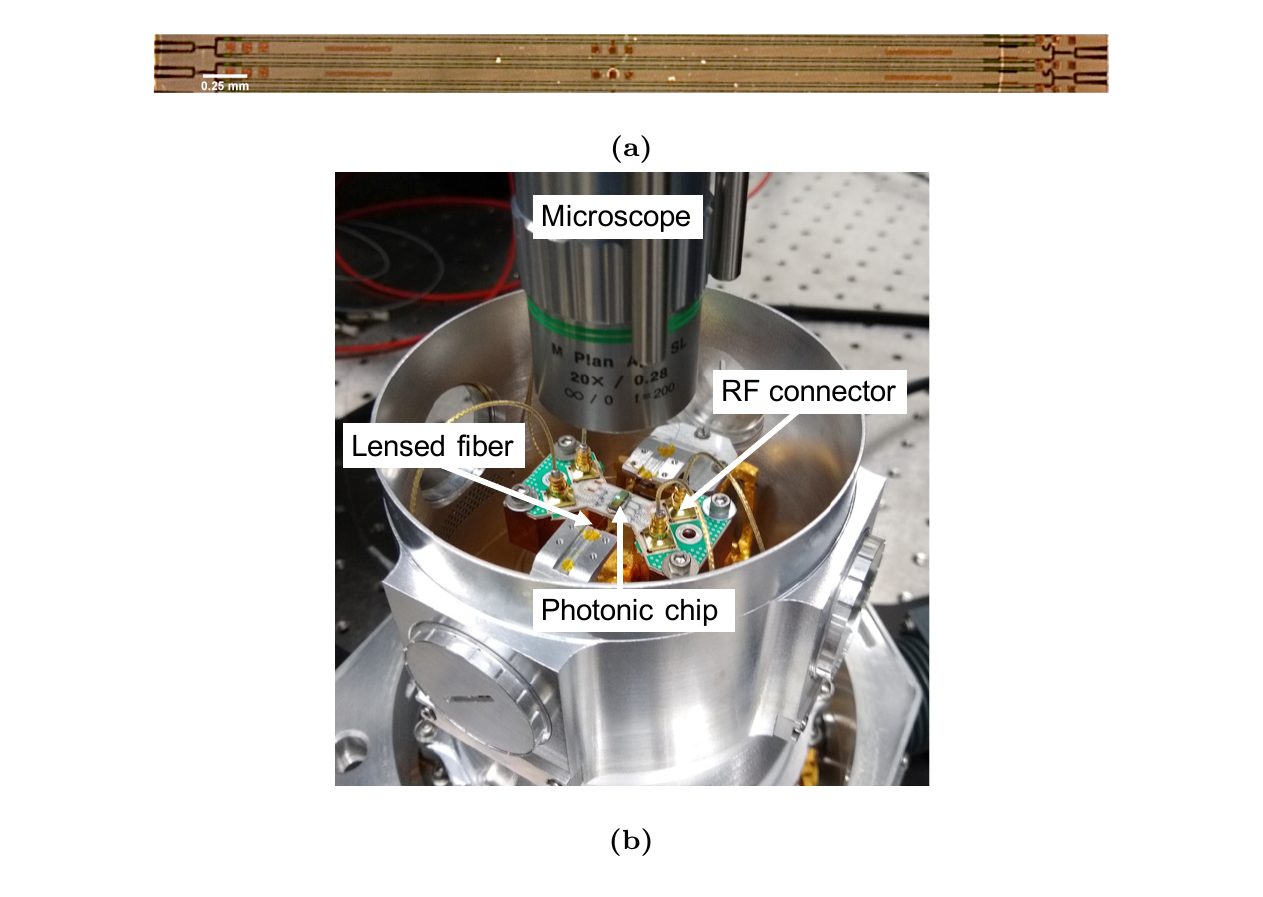}
\caption{\small \textbf{Experimental setup}.
\textbf{(a)} Optical micrograph of the two Mach-Zehnder devices studied in this work. Light is coupled in and out of the devices via the edge couplers at the chip facets. \textbf{(b)} Top view of the cryostat, with center-mounted chip and optical and electrical access.} 
\label{fig:setup}
\end{figure}

Fig.~\ref{fig:setup}a shows the section of the chip containing the Mach-Zehnder devices tested in this work. The packaged chip is shown mounted in the cryostat in Fig.~\ref{fig:setup}b. Wire bonds connect the electrical contact pads on the chip to the PCB's shielded RF signal lines, which have SMP connector terminations. Mini-SMP cables (5 GHz bandwidth) connect the PCB to the cryostat's RF ports for electrical control, and lensed fibers are edge-coupled to the chip facets for optical measurements.

\section{Simulation of effective index shifts} 
\label{sec:appendix-simu}
\setcounter{figure}{0}
The overall effective index shift can be obtained as the sum of the index shifts due to the DC Kerr effect and the plasma dispersion effect. We used the Lumerical CHARGE transport solver to simulate the plasma dispersion effect in our devices according to the Soref-Bennett model \cite{soref1987electrooptical}. Owing to the unavailability of exact information on the dopant implantation parameters for the devices, we approximated the doped regions as having Pearson IV doping profiles with representative parameters based on standard dopant implantation processes \cite{Cheung:2010}. The DC Kerr effect was simulated using the Lumerical FEEM waveguide solver, with the DC-Kerr-induced shift in the refractive index of silicon given by Equation (4) in the main body of the paper, and the third-order nonlinear susceptibility $\chi^{(3)}$ = 2.45 $\times$ 10$^{-19}$~m$^{2}$V$^{-2}$ (as in \cite{Timurdogan:2017jg}). 
The effective index shifts are plotted as a function of increasing reverse bias in Fig.~\ref{fig:simulations}. These simulations provide a qualitative, order-of-magnitude confirmation of the nature of the experimentally-observed index shifts. (Since the exact magnitudes of the index shifts are highly sensitive to the implantation profile, it is not possible to obtain a more exact comparison without knowing the actual dopant implantation profiles of the devices.) While the plasma dispersion-induced index shift is expected to exhibit a dependence close to the square root of the reverse bias voltage, the DC-Kerr-induced index shift is expected to increase quadratically with voltage, producing a total index shift that increases pseudo-linearly with voltage.

\begin{figure*}
\centering
\includegraphics[width=\linewidth]{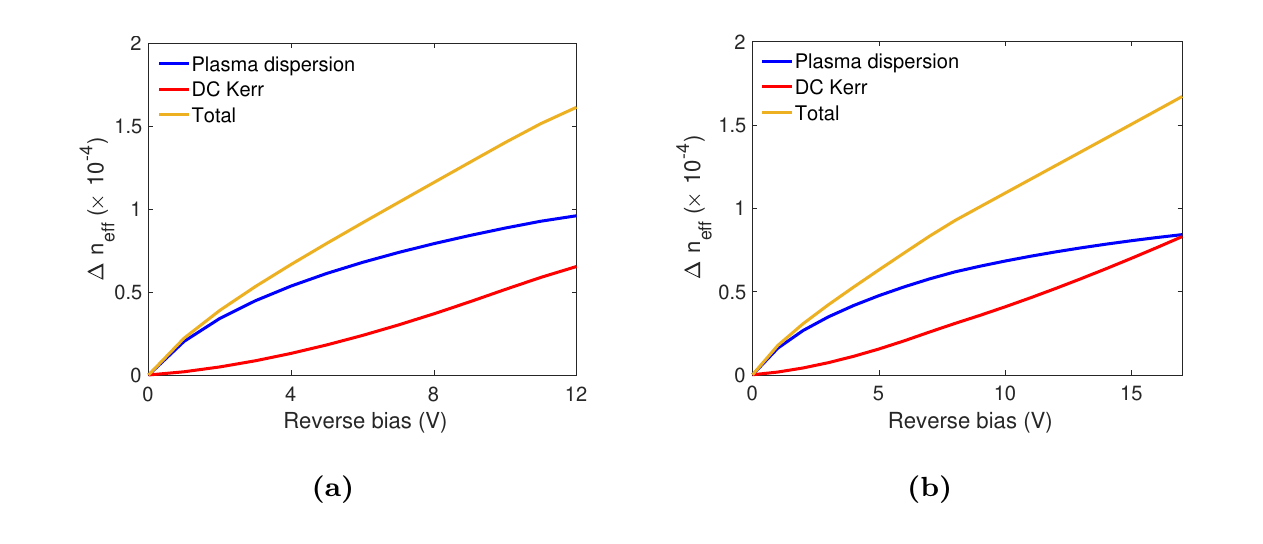} 
\caption{\small \textbf{Plasma dispersion and DC Kerr effects simulated separately and summed for each device with non-uniform doping profiles.} \textbf{(a)} Simulated index shifts for the 300 nm device.   \textbf{(b)} Simulated index shifts for the 400 nm device.}
\label{fig:simulations}
\end{figure*}

\end{document}